\documentclass[preprint,12pt]{elsarticle}

\usepackage{graphics}

\usepackage{amssymb}

\journal{Journal of Crystal Growth}

\begin{document}

\begin{frontmatter}



\title{Volume term of work of critical nucleus formation in terms of chemical potential difference relative to equilibrium one}


\author{Atsushi Mori\footnote{Corresponding author: E-mail atsushimori@tokushima-u.ac.jp, Tel +81-88-656-9417, Fax +81-88-656-9435}}

\address{Institute of Technology and Science, The University of Tokushima,
2-1 Minamijosanjima, Tokushima 770-8506, Japan}

\begin{abstract}
The work of formation of a critical nucleus is sometimes written as $W = n\Delta\mu + \gamma A$.
The first term $W_\mathit{vol} = n\Delta\mu$ is called the volume term and the second term $\gamma A$ the surface term with $\gamma$ being the interfacial tension and $A$ the area of the nucleus.
Nishioka and Kusaka [J.~Chem.~Phys. \textbf{96} (1992) 5370] derived $W_\mathit{vol} = n\Delta\mu$ with $n = V_\beta/v_\beta$ and $\Delta\mu = \mu_\beta(T,p_\alpha) - \mu_\alpha(T,p_\alpha)$ by rewriting $W_\mathit{vol} = -(p_\beta-p_\alpha)V_\beta$ by integrating the isothermal Gibbs-Duhem relation for an incompressible $\beta$ phase, where $\alpha$ and $\beta$ represent the parent and nucleating phases, $V_\beta$ is the volume of the nucleus, $v_\beta$, which is constant, the molecular volume of the $\beta$ phase, $\mu$, $T$, and $p$ denote the chemical potential, the temperature, and the pressure, respectively.
We note here that $\Delta\mu = \mu_\beta(T,p_\alpha) - \mu_\alpha(T,p_\alpha)$ is, in general, not a directly measurable quantity.
In this paper, we have rewritten $W_\mathit{vol} = -(p_\beta-p_\alpha)V_\beta$ in terms of $\mu_\mathit{re} - \mu_\mathit{eq}$, where $\mu_\mathit{re}$ and $\mu_\mathit{eq}$ are the chemical potential of the reservoir (equaling that of the real system, common to the $\alpha$ and $\beta$ phases) and that at equilibrium.
Here, the quantity $\mu_\mathit{re} - \mu_\mathit{eq}$ is the directly measurable supersaturation.
The obtained form is similar to but slightly different from $W_\mathit{vol} = n\Delta\mu$.
\end{abstract}


\end{frontmatter}

\newpage


\section{Introduction \label{sec:intro}}

To calculate the reversible work of formation of a critical nucleus is one of the purposes of the theory of nucleation, because one can predict the steady state nucleation rate $J_{s} = J_0 \exp(-W^*/k_BT)$ through the work of formation of the critical nucleus $W$, where $k_BT$ is the temperature multiplied by Boltzmann's constant.
Here, $W^* \equiv W(R^*)$ is the height of the nucleation barrier with $R^*$ being the radius of the critical nucleus.
We can refer a theory not including molecular level quantities to as a classical nucleation theory. 
We often encounter the following formula [Eq.~(\ref{eq:common})] or equivalent one in the classical nucleation theory: 
\begin{equation}
\label{eq:common}
W = n\Delta\mu + \gamma A,
\end{equation}
with $\gamma$ begin the interfacial tension, $A \equiv 4\pi R^2$ the area of the interface (rigorously speaking, $R$ is the radius of the surface of tension) in textbooks such as \cite{wunderlich,chernov,mutaftschiev,saito,kashchiev,markov} as well as research papers such as \cite{auer2001,gasser2001,kashchiev2004,merikanto2007,wedekind2008,kashchiev2010,kawasaki2010}.
Nishioka and Kusaka~\cite{nishioka1992} called this formula ^^ ^^ a commonly used formula."
In this formula, one regards $\Delta\mu$ the chemical potential difference between the parent phase (the $\alpha$ phase) and the nucleating phase (the $\beta$ phase), i.e., $\Delta\mu \equiv \mu_\beta - \mu_\alpha (<0)$ [in the field of crystal growth, one sometimes defines the supersaturation as $\Delta\mu \equiv \mu_\alpha - \mu_\beta (>0)$ and the negative sign arises as $-n\Delta\mu$].
We note here that the term ^^ ^^ supersaturation" is sometimes used to express the such thermodynamic driving force, apart form the literal meaning.
In this paper, we will adopt this terminology.
One can understand $n$ as the numbers of molecules undergone the phase transition from the $\alpha$ phase to the $\beta$ phase.
Apart form its implication, meaning of $\Delta\mu$ in this expression is immediately ambiguous; for the critical nucleus, the chemical potential of the $\beta$ phase is equal to that of the $\alpha$ phase.
In this paper, we will give correct formulas for $W$ in terms of the chemical potential difference relative to equilibrium one.

The exactly correct form for $W$ given by Gibbs \cite{gibbs} is
\begin{equation}
\label{eq:exact}
W = -(p_\beta - p_\alpha)V_\beta + \gamma A,
\end{equation}
where $p$ denotes the pressure and $V_\beta \equiv 4\pi R^3/3$ is the volume of the nucleus.
Rigorously speaking, $p_\beta$ is the pressure of the hypothetical cluster defined such as possessing the bulk property and filling the inside of the surface of tension, and thus $V_\beta$ the volume inside the surface of tension.
A transparent explanation for the volume term $W_\mathit{vol} = -(p_\beta - p_\alpha)V_\beta$ can be given through a grand potential formalism --- the grand potential is defined as $\Omega = -pV$; the grand potential formalism for interfaces was given in, for example, a textbook by Landau and Lifshitz \cite{landau} and exactness of $W_\mathit{vol} = -(p_\beta - p_\alpha)V_\beta$ is shown by a textbook by Vehkam\"aki \cite{vehkamaki} (see also \cite{yang1983}).
The author wish to introduce a heuristic paper \cite{mori2013} for readers' convenience.
That is, one can readily understand the form of $W_\mathit{vol}$ on the basis of the fact that the reversible work of formation of the critical nucleus is the grand potential difference.
In the following sense, Eq.~(\ref{eq:exact}) is entirely exact.
We divide the process of nucleus formation into two.
One is the formation of a hypothetical cluster of radius $R$, within which the bulk $\beta$ phase fulfills.
The other is the formation of interfacial structure on the mathematical boundary of radius $R$.
The former is given by the first term in Eq.~(\ref{eq:exact}) and the latter is expressed by the second term.
The problem of classical theories is that the interfacial tension $\gamma$ is treated as constant (in particular, in a capillary approximation $\gamma$ for the flat interface is employed), whereas the curvature dependence is unnecessarily negligible \cite{tolman1949,koenig1950,buff1950,bogdan1997,koga1998,bartell2001,blokhuis2006,troster2012}.

In Sec.~\ref{sec:common} we will give the derivation of $W_\mathit{vol} = n [\mu_\beta(T,p_\alpha)-\mu_\alpha(T,p_\alpha)]$ due to Nishioka and Kusaka \cite{nishioka1992}.
Here, we will give a brief review on this form.
To express the pressures in terms of the chemical potentials using the Gibbs-Duhem relation was seen in Oxtoby and Kashchiev's paper on the nucleation theorem \cite{oxtoby1994}.
In their paper, however, $W_\mathit{vol} $ was not calculated.
Laaksonen~\textit{et~al.} \cite{laaksonen1999}  pointed out that $W_\mathit{vol}$ could be calculated by extending the Oxtoby and Kashchiev's line.
This is completely the same as Nishioka and Kusaka's \cite{nishioka1992}.
Also, the same procedure was followed by Debenedetti and Reiss \cite{debenedetti1998}.
In order to avoid defocusing of the point of the present paper, we will not give a general review on the nucleation theorem.

Differentiating Eq.~(\ref{eq:exact}) with respect to $R$ and equating zero and then solving for $R$, we have formula for the size of the critical nucleus as
\begin{equation}
\label{eq:criticalR}
R^* = \frac{2\gamma}{p_\beta-p_\alpha}.
\end{equation}
By substituting Eq.~(\ref{eq:criticalR}) for $R$ in Eq.~(\ref{eq:exact}) the hight of the nucleation barrier is obtained as
\begin{equation}
\label{eq:criticalW}
W^* = \frac{1}{2}(p_\beta-p_\alpha)V_\beta^* = \frac{1}{3}\gamma A^* = \frac{16\pi\gamma^3}{3(p_\beta-p_\alpha)^2},
\end{equation}
with $V_\beta^* \equiv 4\pi (R^*)^3/3$ and $A^* \equiv 4\pi (R^*)^2$.
Corresponding to Eq.~(\ref{eq:common}), one has
\begin{equation}
W^* = \frac{16\pi v^2 \gamma^3}{3(\Delta\mu)^2},
\end{equation}
instead of the last expression of Eq.~(\ref{eq:criticalW}).
Here, $v$ represent ^^ ^^ the molecular volume"; that is, $n$ (or $n^*$) is expressed as $V_\beta/v$ (or $V_\beta^*/v$).
We note here that the definition of $v$ is immediately ambiguous; whose phase is not specified, or $v$ may be common to the $\alpha$ and $\beta$ phases.
Also, as mentioned above, the definition of $\Delta\mu$ immediately unclear.
For a rarefied $\alpha$ phase, one uses very often
\begin{equation}
\label{eq:lnS}
\Delta\mu = k_BT \ln S,
\end{equation}
where $S$ is the supersaturation ratio $p_\mathit{re}/p_\mathit{eq}$ with  $p_\mathit{re}$ being the pressure of the reservoir, which equals $p_\alpha$, and $p_\mathit{eq}$ the equilibrium pressure.
In other words, $\Delta\mu$ is defined as
\begin{equation}
\label{eq:incorrect}
\Delta\mu = \mu_\alpha(T, p_\mathit{re}) - \mu_\alpha(T, p_\mathit{eq}),
\end{equation}
where $T$ is the temperature, which is assumed to be uniform throughout the system.
Hereafter, $T$ will be omitted for brevity.
As shall be shown below, this definition is, however, incorrect.
Nevertheless, expression $W^*$ in terms of this $\Delta\mu$ is strongly desired because this quantity is directly measurable. 
Indeed, Eq.~(\ref{eq:incorrect}) is widely used such as in \cite{mcgraw1981,wilemski1991,yasuoka1998,holten2005,noguera2006,wasai2007,uwaha2010}.

\section{Derivation of commonly used formula \label{sec:common}}

Nishioka and Kusaka \cite{nishioka1992} found out that in case that the $\beta$ phase is incompressible, the volume term $W_\mathit{vol} = -(p_\beta - p_\alpha)V_\beta$ can be rewritten in the form $n(\mu_\beta-\mu_\alpha)$.
That is, they integrated
\begin{equation}
\label{eq:dmudp}
\left(\frac{\partial\mu}{\partial p}\right)_T = v,
\end{equation} 
which is nothing other than Gibbs-Duhem relation for the isothermal case, for the $\beta$ phase.
Unfortunately, they concluded incorrectly that the form (\ref{eq:common}) was valid only for the case of the incompressible $\beta$ phase such as the nucleation of an incompressible liquid phase in a vapor phase.
For example, this condition is, in a mathematical form, valid for a bubble nucleation in an incompressible liquid phase.
In this paper, however, we limit ourselves to the incompressible $\beta$ phase to avoid the confusion in argument.
Note that the form of Eq.~(\ref{eq:common}) cannot be derived in general --- we can derive in some approximations.

\begin{figure*}[tb]
\includegraphics{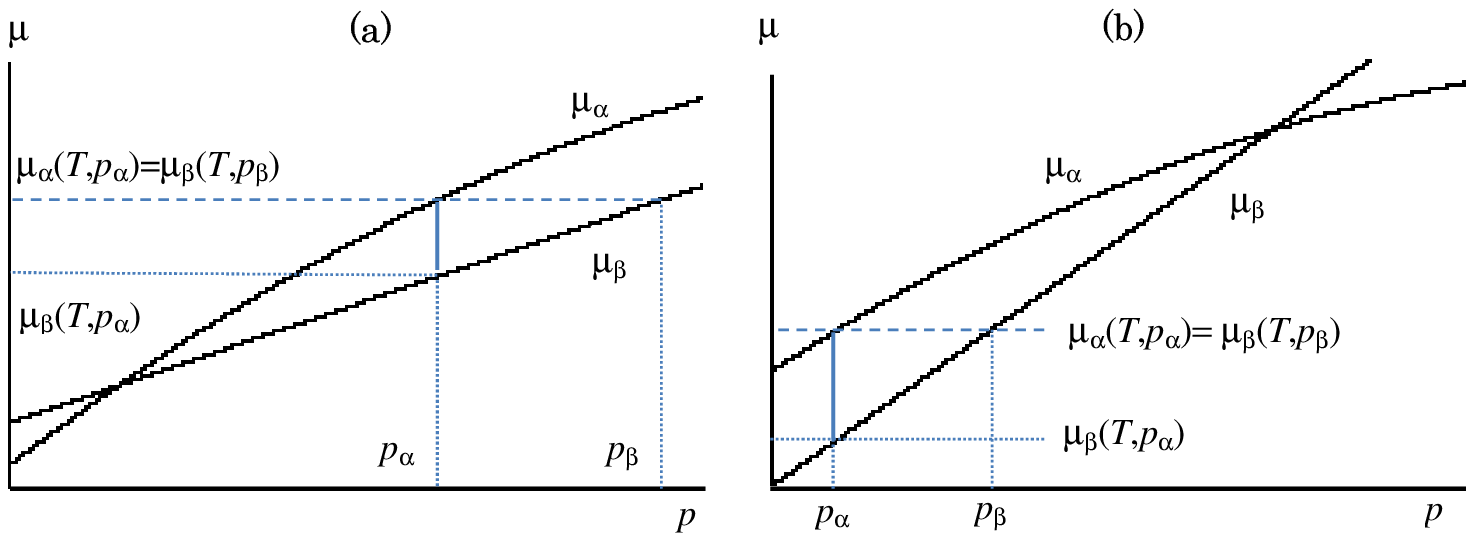}
\caption{\label{fig:mudrop} $\mu$-$p$ relation for cases of incompressible $\beta$ phases; (a) a normal case that the $\beta$ phase is denser than the $\alpha$ phase and (b) an abnormal case that the $\alpha$ phase is denser than the $\beta$ phase. 
The horizontal dashed lines indicate the chemical potential, which is common to the nucleus and the parent phase.
The vertical solid lines depict $\mu_\beta(p_\alpha) - \mu_\alpha(p_\alpha)$ [eq.~(\ref{eq:wvolbeta})].}
\end{figure*}

For the case of the incompressible $\beta$ phase, $W_\mathit{vol}$ is given by \cite{nishioka1992}
\begin{equation}
\label{eq:wvolbeta}
W_\mathit{vol} =\frac{V_\beta}{v_\beta} [\mu_\beta(p_\alpha) - \mu_\alpha(p_\alpha)].
\end{equation}
To reach to this expression we have used the fact that the chemical potential of the nucleation phase $\mu_\beta(p_\beta)$ is equal to that of the parent phase $\mu_\alpha(p_\alpha)$, i.e.,
\begin{equation}
\mu_\beta(p_\beta) = \mu_\alpha(p_\alpha),
\end{equation}
 (that is, the chemical potential is uniform throughout the system) as indicated by a horizontal dashed lines in Fig.~\ref{fig:mudrop}.
The vertical solid lines in Fig.~\ref{fig:mudrop} depicts $\mu_\beta(p_\alpha) - \mu_\alpha(p_\alpha)$ [Eq.~(\ref{eq:wvolbeta})].
Figure~\ref{fig:mudrop}~(a) is for a normal case such as a liquid droplet in a vapor phase.
On the other hand, Fig.~\ref{fig:mudrop}~(b) is for an abnormal case such as formation a nucleus of less denser, incompressible $\beta$ phase in a denser $\alpha$ phase.
In both cases, $\mu_\beta(p_\alpha) - \mu_\alpha(p_\alpha)$ does not coincide to the supersaturation, which is the chemical potential difference relative to the chemical potential at the $\alpha$-$\beta$ phase equilibrium (sometimes with a negative sign as note in Sec.~\ref{sec:intro}).

\section{Volume term in general \label{sec:general}}

\begin{figure*}[tb]
\includegraphics{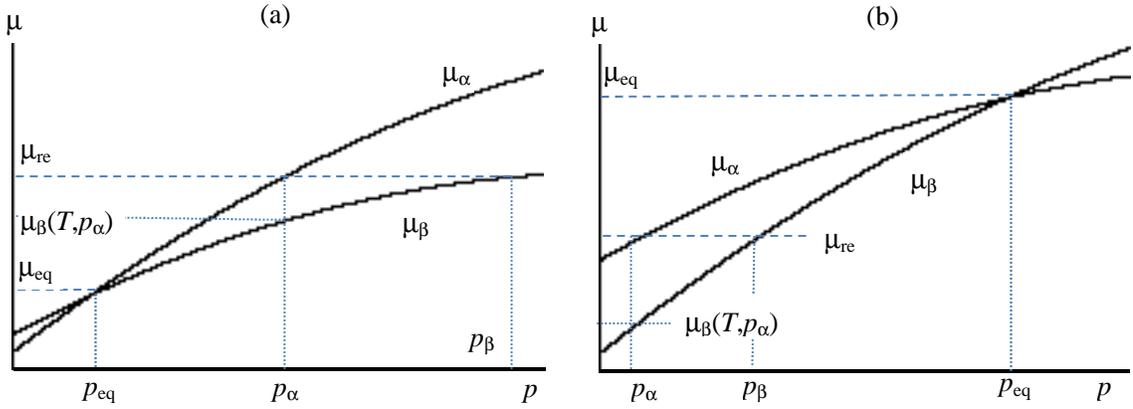}
\caption{\label{fig:mugeneral} Location of $\mu_\mathit{eq}$, $p_\beta$, $p_\alpha$, and $p_\mathit{eq}$ is illustrated (note that the reserver pressure $p_\mathit{re}$ coincides to $p_\alpha$); (a) a case of condensation, i.e., the $\beta$ phase is denser than the $\alpha$ phase, and (b) a case of bubble nucleation, i.e., the $\alpha$ phase is denser than the $\beta$ phase.
In case (a), $p_\beta>p_\alpha>p_\mathit{eq}$ holds.
On the other hand, in case (b), $p_\alpha<p_\beta<p_\mathit{eq}$ holds.}
\end{figure*}

In these ways, one knows two cases where the form of eq.~(\ref{eq:common}) is valid.
Let us develop a general consideration; we consider the case that $v$ in eq.~(\ref{eq:dmudp}) is a function of $p$ according to the mean value theorem.
Following Nishioka and Kusaka \cite{nishioka1992} we integrate the Gibbs-Duhem relation of the form of eq.~(\ref{eq:dmudp}) from $p_1$ to $p_2$.
\begin{equation}
\label{eq:tmv}
\mu(p_2) - \mu(p_1) = v(\tilde{p})(p_2-p_1),
\end{equation}
where $\tilde{p}$ is a certain value lying in the interval ($p_2$,$p_1$).
Let us define $p_\mathit{eq}$ as the equilibrium pressure of the $\alpha$-$\beta$ phase equilibrium (the saturation pressure of the $\alpha$ phase with respect to the $\beta$ phase), which is the solution to
\begin{equation}
\mu_\beta(p)=\mu_\alpha(p).
\end{equation}
For latter convenience, let us define an abbreviation $\mu_\mathit{eq}$ for $\mu_\beta(p_\mathit{eq}) = \mu_\alpha(p_\mathit{eq})$.
Also for latter convenience, we define $\mu_\mathit{re} \equiv \mu_\beta(p_\beta) = \mu_\alpha(p_\alpha)$, which is the chemical potential of the reservoir in $\mu VT$ ensemble.
Applying eq.~(\ref{eq:tmv}) for the $\beta$ and $\alpha$ phase, we have $\mu_\beta(p_\beta) - \mu_\beta(p_\mathit{eq}) = v_\beta(p_\mathit{eq}+\theta_\beta(p_\beta-p_\mathit{eq}))(p_\beta-p_\mathit{eq})$ and $\mu_\alpha(p_\alpha) - \mu_\alpha(p_\mathit{eq}) = v_\alpha(p_\mathit{eq}+\theta_\alpha(p_\alpha-p_\mathit{eq}))(p_\alpha-p_\mathit{eq})$ with $\theta_\beta$ and $\theta_\alpha$ being certain values laying in an interval (0,1), i.e.,
\begin{eqnarray}
\label{eq:tmvbeta}
\mu_\mathit{re} - \mu_\mathit{eq} &=& v_\beta(\tilde{p_\beta})(p_\beta-p_\mathit{eq}), \\
\label{eq:tmvalpha}
\mu_\mathit{re} - \mu_\mathit{eq} &=& v_\alpha(\tilde{p_\alpha})(p_\alpha-p_\mathit{eq}),
\end{eqnarray}
where $v_\beta$ and $v_\alpha$ are molecular volumes of respective phase, $\tilde{p_\beta}$ and $\tilde{p_\alpha}$ certain values respectively lying in intervals ($p_\mathit{eq}$,$p_\beta$) and ($p_\mathit{eq}$,$p_\alpha$) for cases of $p_\beta > p_\alpha > p_\mathit{eq}$ as shown in Fig.~\ref{fig:mugeneral}~(a) [for cases of $p_\alpha < p_\beta < p_\mathit{eq}$ as shown in Fig.~\ref{fig:mugeneral}~(b) the intervals are replaced with ($p_\beta$, $p_\mathit{eq}$) and ($p_\alpha$, $p_\mathit{eq}$), respectively].
By dividing Eq.~(\ref{eq:tmvbeta}) by $v_\beta(\tilde{p_\beta})$ and Eq.~(\ref{eq:tmvalpha}) by $v_\alpha(\tilde{p_\alpha})$ and subtracting the latter from the former, we eliminate $p_\mathit{eq}$ to have
\begin{equation}
\label{eq:diffp}
\left[\frac{1}{v_\beta(\tilde{p_\beta})}-\frac{1}{v_\alpha(\tilde{p_\alpha})}\right](\mu_\mathit{re}-\mu_\mathit{eq}) = p_\beta-p_\alpha.
\end{equation}
We can rewrite the first term in Eq.~(\ref{eq:exact}) by substituting $p_\beta-p_\alpha$ by Eq.~(\ref{eq:diffp}).
\begin{eqnarray}
\nonumber
W_\mathit{vol} &=& -\left[\frac{1}{v_\beta(\tilde{p_\beta})}-\frac{1}{v_\alpha(\tilde{p_\alpha})}\right](\mu_\mathit{re}-\mu_\mathit{eq}) V_\beta \\
\label{eq:wvoltmv}
&=& -(n_\beta-n_\alpha)(\mu_\mathit{re}-\mu_\mathit{eq}),
\end{eqnarray}
where $n_\beta \equiv V_\beta/v_\beta(\tilde{p_\beta})$ and $n_\alpha \equiv V_\beta/v_\alpha(\tilde{p_\alpha})$.
This form is slightly different from the first term in Eq.~(\ref{eq:common}).
In a case that the $\alpha$ phase is a rarefied vapor phase, the quantity $n_\alpha$ tends to vanish and the form of the first term in eq.~(\ref{eq:common}) is obtained.
This case can be included in the case of Fig.~\ref{fig:mudrop}~(a) (nucleation of an incompressible liquid droplet in an infinitely rarefied vapor).
In this case the curve of $\mu_\alpha$ tends to a vertical line.
In this limiting case the horizontal location of the intersection of horizontal dashed line with the $\mu_\alpha$ curve locates on the intersection of two $\mu$ curves; that is, $\mu_\beta(p_\alpha)-\mu_\alpha(p_\alpha)$ in eq.~(\ref{eq:wvolbeta}) tends to coincide to $\mu_\mathit{re}-\mu_\mathit{eq}$.
Here, we note that $-(\mu_\mathit{re}-\mu_\mathit{eq})$ is the true supersaturation for the case of $p_\beta > p_\alpha > p_\mathit{eq}$ [Fig.~\ref{fig:mugeneral}~(a)] and the true undersaturation for the case of $p_\beta < p_\alpha < p_\mathit{eq}$ [Fig.~\ref{fig:mugeneral}~(b)].
In this way, we have successfully express the work term in terms of the supersaturation and revealed under what case the commonly used formula holds.

We have obtained a form of Eq.~(\ref{eq:common}) with $\Delta\mu = \mu_\mathit{re} - \mu_\mathit{eq}$.
The result of Eq.~(\ref{eq:wvoltmv}) includes an issue concerning physicochemical problems, although it is of a mathematically beautiful form; in definitions $n_\beta \equiv V_\beta/v_\beta(\tilde{p_\beta})$ and $n_\alpha \equiv V_\beta/v_\alpha(\tilde{p_\alpha})$, the denominators $v_\beta(\tilde{p_\beta})$ and $v_\alpha(\tilde{p_\alpha})$ are not fixed constant values.
We will give a prompt solution here.
Assuming a smallness of $\mu_\mathit{re} - \mu_\mathit{eq}$ [rigorously speaking the smallness should be described in terms of a dimensionless quantity --- that is, the present statement reads $|(\mu_\mathit{re} - \mu_\mathit{eq})/\mu_\mathit{eq}| \ll 1$ or $|\mu_\mathit{re} - \mu_\mathit{eq}|/k_BT \ll 1$], let us make an second-order expansion instead of Eqs.~(\ref{eq:tmvbeta}) and (\ref{eq:tmvalpha}).
Instead of Eq.~(\ref{eq:dmudp}), it is more convenient to start with
\begin{equation}
\label{eq:dpdmu}
\left(\frac{\partial p}{\partial\mu}\right)_T = \rho,
\end{equation} 
with $\rho=1/v$ being the number density.
The second-order expansions are
\begin{eqnarray}
\nonumber
p_\beta-p_\mathit{eq} &=&
\rho_\beta(p_\mathit{eq}) (\mu_\mathit{re} - \mu_\mathit{eq})
+ \frac{1}{2} \rho_\beta^2(p_\mathit{eq}) \kappa_\beta(p_\mathit{eq}) (\mu_\mathit{re} - \mu_\mathit{eq})^2 \\
\label{eq:tpbetapeq}
&&
+ O((\mu_\mathit{re} - \mu_\mathit{eq})^3), \\
\nonumber
p_\alpha-p_\mathit{eq}  &=&
\rho_\alpha(p_\mathit{eq}) (\mu_\mathit{re} - \mu_\mathit{eq})
+ \frac{1}{2} \rho_\alpha^2(p_\mathit{eq}) \kappa_\alpha(p_\mathit{eq}) (\mu_\mathit{re} - \mu_\mathit{eq})^2 \\
\label{eq:palphaeq}
&&
+ O((\mu_\mathit{re} - \mu_\mathit{eq})^3),
\end{eqnarray}
where $\kappa$ denotes the isothermal compressibility.
Subtraction Eq.~(\ref{eq:palphaeq}) from Eq.~(\ref{eq:tpbetapeq}), we have
\begin{eqnarray}
\nonumber
p_\beta-p_\alpha &=&
[\rho_\beta(p_\mathit{eq})-\rho_\alpha(p_\mathit{eq})] (\mu_\mathit{re} - \mu_\mathit{eq}) \\
\nonumber
&& + \frac{1}{2} [\rho_\beta^2(p_\mathit{eq}) \kappa_\beta(p_\mathit{eq}) - \rho_\alpha^2(p_\mathit{eq}) \kappa_\alpha(p_\mathit{eq})]
(\mu_\mathit{re} - \mu_\mathit{eq})^2 \\
\label{eq:pbetapalba2nd}
&&
+ O((\mu_\mathit{re} - \mu_\mathit{eq})^3).
\end{eqnarray}
Neglecting the second and higher order terms in Eq.~(\ref{eq:pbetapalba2nd}) and inserting in the first term of Eq.~(\ref{eq:exact}), we have the form of Eq.~(\ref{eq:wvoltmv}) with fixed $n_\beta$ and $n_\alpha$ as an approximation.
Taking into account the second term, we can improve the approximation.

Coincidence between the present result and the commonly used formula with Eq.~(\ref{eq:incorrect}) can be understood as follows.
For nucleations of incompressible $\beta$ phase in a rarefied gas $\mu_\beta(p_\alpha) - \mu_\alpha(p_\alpha)$ can be rewritten as
\begin{eqnarray}
\nonumber
\mu_\beta(p_\alpha) - \mu_\alpha(p_\alpha) &=& [\mu_\beta(p_\alpha)-\mu_\mathit{eq}] - [\mu_\alpha(p_\alpha)-\mu_\mathit{eq}] \\
\nonumber
&=& [\mu_\beta(p_\alpha)-\mu_\beta(p_\mathit{eq})] - [\mu_\alpha(p_\alpha)-\mu_\alpha(p_\mathit{eq})] \\
\label{eq:kblnS}
&=& v_\beta (p_\alpha-p_\mathit{eq}) - k_BT \ln \frac{p_\alpha}{p_\mathit{eq}}.
\end{eqnarray}
The last term can be expanded in $p_\alpha-p_\mathit{eq}$ as
\begin{eqnarray}
\nonumber
k_BT \ln \frac{p_\alpha}{p_\mathit{eq}} &=& k_BT \ln \left[ 1 + \frac{p_\alpha - p_\mathit{eq}}{p_\mathit{eq}} \right] \\
\nonumber
&\cong& k_BT \frac{p_\alpha - p_\mathit{eq}}{p_\mathit{eq}} \\
&=& v_\alpha(p_\mathit{eq}) (p_\alpha - p_\mathit{eq}).
\end{eqnarray}
To reach to the last line, the equation of state for the ideal gas has been used.
Because the $\alpha$ phase is a rarefied gas, the inequality $v_\alpha \gg v_\beta$ holds and then we find that the first term in Eq.~(\ref{eq:kblnS}) can be neglected.

\section{Discussion \label{sec:discussion}}

In 1984, Wilemski \cite{wilemski1984} divided the number of molecule of species $i$ included in a nucleus for the binary system as
\begin{equation}
\label{eq:nbns}
n_i = n_i^b + n_i^s,
\end{equation}
with the superscripts $b$ and $s$ denoting bulk and surface and wrote down the condition of the critical nucleus as
\begin{eqnarray}
\nonumber
0 &=& (\Delta\mu + \gamma(\partial A/n_j)_{n_i})(dn_j)_{n_i} \\
\nonumber
&& + n_1^b d\mu_1^l + n_2^b d\mu_2^l \\
\label{eq:wilemski}
&&+ n_1^s d\mu_1^l + n_2^s d\mu_2^l + Ad\gamma \mbox{\hspace{1em} (const. $T$,$P$)}.
\end{eqnarray}
In 1999, Laaksonen~\textit{et~al.} \cite{laaksonen1999} revealed the work of formation of the nucleus underlaying Eq.~(\ref{eq:wilemski}) as
\begin{equation}
\label{eq:gdeltamu}
\Delta G = \sum_i (\mu_\mathit{li}(P_v) - \mu_\mathit{vi}(P_v)) g_i + A\gamma,
\end{equation}
with $g_i \equiv n_\mathit{li} - n_\mathit{vi} + n_\mathit{si}$,  where the subscripts l and v denote the liquid and vapor phases (in Wilemski's paper, the starting equation is $\Delta G = n_1 \Delta\mu_1 + n_2 \Delta\mu_2 + A\gamma$).
The so-called surface excess number of molecules, $n_s$, is identical to the superficial number of molecules in the Gibbs interfacial thermodynamics \cite{gibbs} and we find that $n_s$ is proportional to $A$.

As mentioned in Sec.~\ref{sec:intro}, in the Gibbs interfacial thermodynamics the work of nucleus formation is divided into the formation of the hypothetical cluster and that of the interfacial structure.
A quantity proportional to a superficial quantity is categorized as the latter.
In this respect, in Eq.~(\ref{eq:gdeltamu}) the volume term is regarded as
\begin{equation}
\label{eq:wvollaaknonen}
W_\mathit{vol} = \sum_i (\mu_\mathit{li}(P_\mathit{v}) - \mu_\mathit{vi}(P_\mathit{v})) (n_\mathit{li} - n_\mathit{vi})
\end{equation}
While the factors regarded as $\Delta\mu$ are different with each other in Eqs.~(\ref{eq:wvoltmv}) and (\ref{eq:wvollaaknonen}),
the coefficients to ^^ ^^ $\Delta\mu$" coincide with each other.
The difference is that in Eq.~(\ref{eq:gdeltamu}) the quantity $\mu_\beta(p_\alpha)-\mu_\alpha(p_\alpha)$ appears, not the quantity $\mu_\mathit{re}-\mu_\mathit{eq}$.

\section{Concluding remarks \label{sec:summary}}

We have successfully rewritten the volume term of the work of formation of a critical nucleus in terms of the supersaturation.
The result is similar to the form of $W_\mathit{vol} = n\Delta\mu$ but slightly different; $n$ in this form has been replaced with $n_\beta-n_\alpha$.
The form $W_\mathit{vol} = n\Delta\mu$ with $\Delta\mu$ being the supersaturation (the chemical potential difference relative to equilibrium) is recovered in a limiting case that the parent phase is a rarefied gas.
This is a finding that requires a concerning note in textbooks --- as mentioned in Sec.~\ref{sec:intro}, some textbooks lead readers to understanding that $W_\mathit{vol} = n\Delta\mu$ is exact, and in some literatures this form is valid only for an incompressible nucleating phase.

We wish to postpone comparisons with experimental studies after formulation in a form of the nucleation theorem.
That is, $W^*$ should be plotted against experimentally determined $\Delta\mu$'s to evaluate the deferential coefficient.
In relation with experiments, a crucial comment arises: the mathematical formula itself (without expansion made in the latter part of Sec.~\ref{sec:general}) makes a sense --- if one forcibly write the work of formation of a critical nucleus in the form like Eq.~(\ref{eq:common}), then uncertainties necessarily accompany in interpreting the experimental results.

\section*{acknowledgment}
The author gratefully acknowledges discussions with Dr.~Y.~Suzuki.
Also he thanks Prof.~E.~Yokoyama for reading the manuscript.








\newpage

\newpage

\end{document}